\documentclass[preprint,12pt]{elsarticle}
\usepackage{bm}

\usepackage{amssymb}
\usepackage{amsmath}

\journal{Journal of Subatomic Particles and Cosmology}

\begin{document}

\begin{frontmatter}



\title{Rapidity scan with DCCI at LHC energy}


\author[1]{Shin-ei Fujii} 
\author[2]{Yasuki Tachibana}
\author[1]{Tetsufumi Hirano}

\affiliation[1]{organization={Department of Physics, Sophia University},
            city={Tokyo},
            postcode={102-8554}, 
            country={Japan}}
\affiliation[2]{organization={Akita International University},
            city={Yuwa, Akita-city},
            postcode={010-1292}, 
            country={Japan}}

\begin{abstract}
We extend the dynamical core-corona initialization (DCCI2) model to include the baryon number evolution in the entire system created in high-energy heavy-ion collisions. 
Introducing the source term for the baryon number, we describe the early-stage equilibration and later-stage hydrodynamic evolution of the baryon number throughout the system from midrapidity to forward rapidity. 
Through numerical simulations with this extended model, we show that extremely large baryon chemical potentials are realized in forward rapidity regions at the LHC energy and are comparable to those of the BES energies. 
Moreover, we show that fluctuations of baryon chemical potentials are large and, consequently, negative baryon chemical potential regions appear at midrapidity.
\end{abstract}



\begin{keyword}
Quark-gluon plasma \sep Relativistic heavy-ion collisions \sep Core-corona picture \sep
Relativistic hydrodynamics \sep Rapidity scan \sep Finite baryon number


\end{keyword}

\end{frontmatter}



\section{Introduction}
\label{sec1}
In recent years, analyses using RHIC-BES data have been actively conducted to explore the high baryon chemical potential region in the QCD phase diagram.
Meanwhile, in higher-energy collisions, such as the top RHIC and LHC energies, the presence of high baryon density in the forward rapidity region has been suggested \cite{Li}.
This implies that, in addition to RHIC-BES, a rapidity scan---analysis in the longitudinal direction---could serve as a complementary way to explore the high baryon chemical potential region in the QCD phase diagram.
However, it is not trivial at all whether the high baryon density in the forward rapidity region is achieved as equilibrium matter. 

The dynamical core-corona initialization (DCCI2) model \cite{Kanakubo} is a state-of-the-art framework that dynamically describes the space-time evolution of both equilibrium and non-equilibrium components, including their interactions.
In this study, we extend the DCCI2 model to describe the evolution of baryon number, including both its dynamical deposition into a locally equilibrated QGP fluid and its subsequent hydrodynamic evolution. 
Using this model, we show the baryon number distribution of core (locally equilibrated fluid) and corona (non-equilibrated particles) component across the entire system at the LHC energy.

\section{Model}
\label{sec2}
In DCCI2, the initial partons are generated using \textsc{Pythia8} \cite{pythia8} and \textsc{Pythia8 Angantyr} \cite{pythia8_angantyr}.
Dynamical fluidization of partons is described by the energy-momentum source term $j^{\nu}$ in the continuity equation:
\begin{equation}
\partial_{\mu} T^{\mu\nu}_{\rm{fluid}} = j^{\nu}, \quad 
j^{\nu} = - \sum^{N_\mathrm{parton}}_{i} \frac{dp^{\nu}_{i}(t)}{dt} G ( \bm{x} - \bm{x}_{i}(t)).
\end{equation}
Here, $j^{\nu}$ is the energy-momentum source term, $p^{\nu}_{i}$ is the four-momentum of the $i$th parton, $G$ is the three-dimensional Gaussian function centered at the position of the $i$th parton $\bm{x}_{i}(t)$. 
The energy loss rate ${dp^{\nu}_{i}(t)}/{dt}$ is calculated from a phenomenological parameterization that depends on the transverse momentum, the density of surrounding partons, and the relative velocity~\cite{Kanakubo}. 
When a parton deposits all energy into the fluid, it is regarded as a dead parton.

We extend the hydrodynamic module adding the baryon number conservation equation with the baryon number source term $\rho$:
\begin{equation}
\partial_{\mu} N^{\mu}_{\rm{fluid}} = \rho,\qquad 
\rho = - \sum^{N_\mathrm{dead}}_{j} \frac{dB_{j}}{dt} G ( \bm{x} - \bm{x}_{j}(t)),
\end{equation}
where $B_{j}$ is the baryon number of the $j$th dead parton.
The summation runs over dead partons at each time step.
This extension enables us to describe the fluidization and hydrodynamic evolution of the baryon number. 
For the equation of state, we employ the NEOS-BQS model~\cite{NEOS} which incorporates finite baryon, charge, and strangeness chemical potentials consistent with the conserved charge susceptibilities from lattice QCD calculations.

\section{Results}
\begin{figure}[h]
  \centering
  \begin{minipage}[t]{0.48\linewidth}
    \centering
    \includegraphics[width=\linewidth,clip]{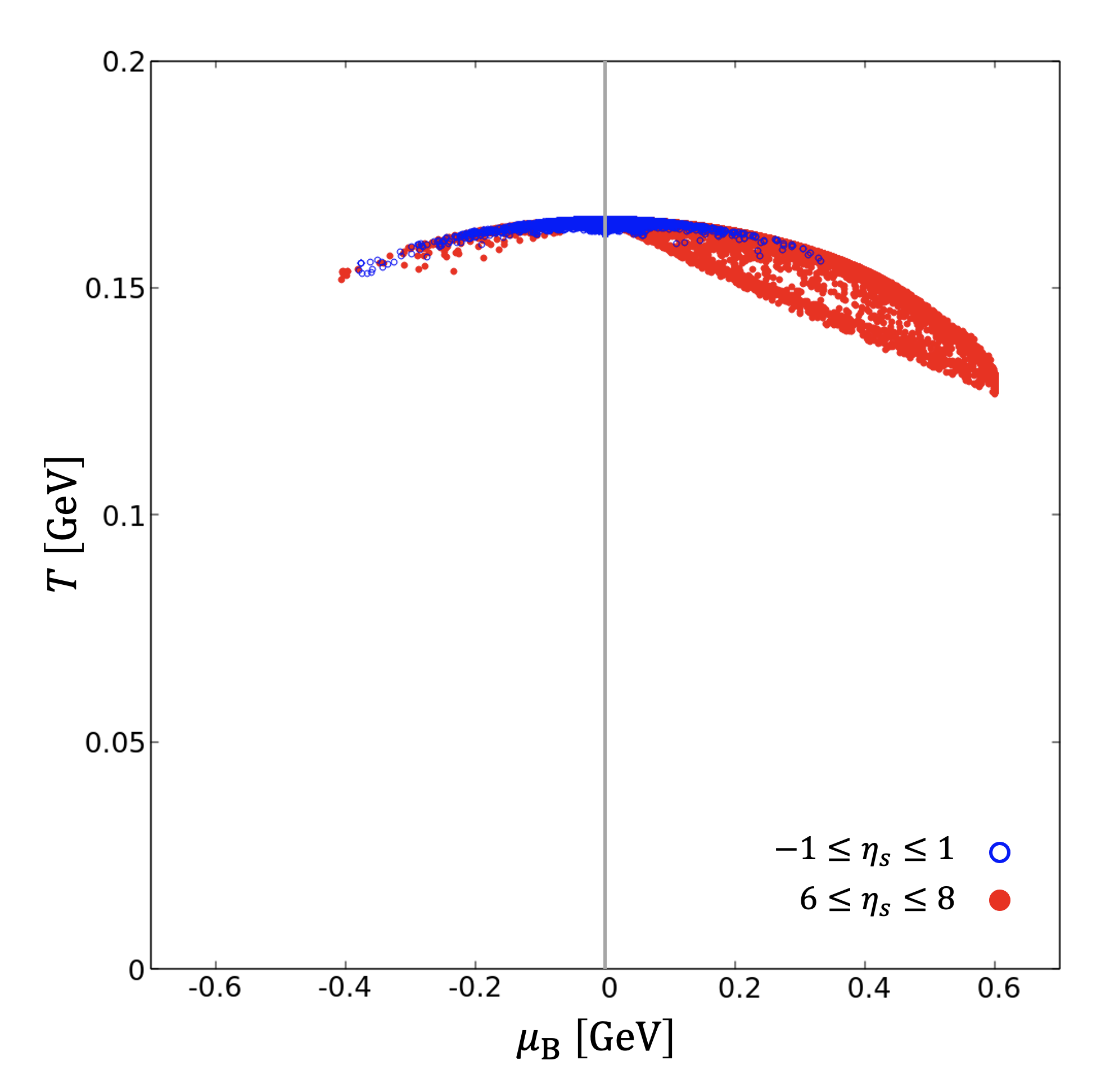}
    \caption{Temperature and baryon chemical potential of each hypersurface element within rapidity ranges $-1 \leqq \eta_{\mathrm{s}} \leqq 1$ and $6 \leqq \eta_{\mathrm{s}} \leqq 8$ on $T$-$\mu_{\mathrm{B}}$ plane.}
    \label{fig:hypersurface1}
  \end{minipage}
  \hspace{0.02\linewidth}
  \begin{minipage}[t]{0.48\linewidth}
    \centering
    \includegraphics[width=\linewidth,clip]{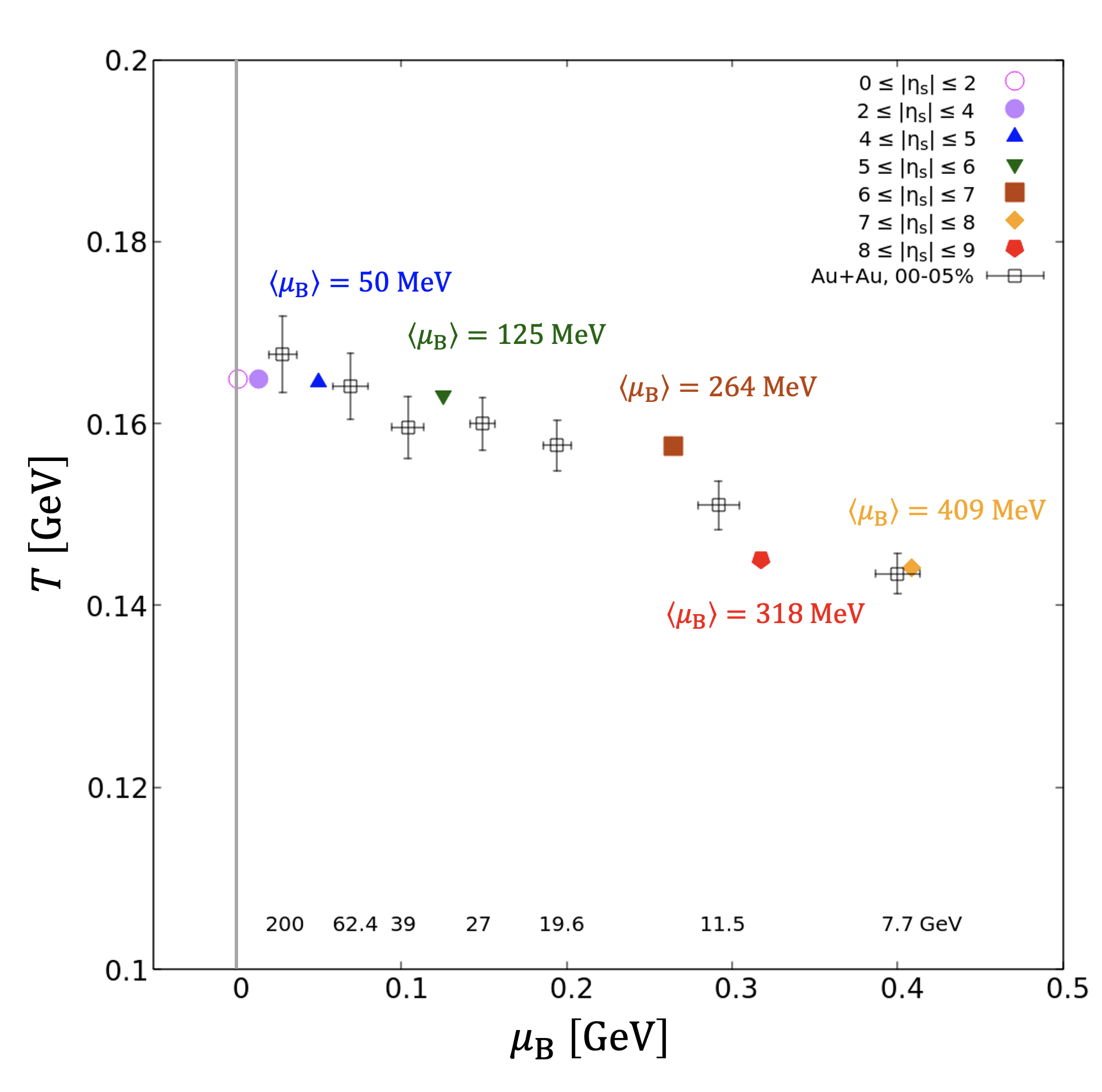}
    \caption{Averaged temperature and baryon chemical potential of  particlization hypersurface elements within specific rapidity ranges.
    Plots with error bars are chemical freezeout parameters obtained from RHIC-BES data \cite{BES}.}
    \label{fig:hypersurface2}
  \end{minipage}
\end{figure}
Throughout this section, we show the results obtained from a single event simulation of a Pb+Pb collision at $\sqrt{s_{\mathrm{NN}}} = 2.76\, \mathrm{TeV}$ with the impact parameter $b = 2.46\, \mathrm{fm}$.
We perform the particlization in the hypersurface at constant energy density $e(T,\mu_{\mathrm{B}}) = 0.547\, \mathrm{GeV/fm^{3}}$.

Figure \ref{fig:hypersurface1} shows temperature and baryon chemical potential of each particlization hypersurface element within the space-time rapidity ranges $-1 \leqq \eta_{\mathrm{s}} \leqq 1$ and $6 \leqq \eta_{\mathrm{s}} \leqq 8$.
Although the value of $\mu_{\mathrm{B}}$ averaged in the midrapidity region is almost vanishing,  its fluctuations have a substantial presence and some hypersurface elements reach $\mu_{\mathrm{B}} \approx -0.4\, \mathrm{GeV}$ in the midrapidity region of $-1 \leqq \eta_{\mathrm{s}} \leqq 1$.
Thus, negative $\mu_{\mathrm{B}}$ should be carefully treated in the sophistication stage of dynamical models since anti-particles would be emitted more than uniformly $\mu_{\mathrm{B}} = 0$ medium at midrapidity.
In forward rapidities, $6 \leqq \eta_{\mathrm{s}} \leqq 8$, the typical $\mu_{\mathrm{B}}$ obviously becomes large and some elements reach $\mu_{\mathrm{B}} = 0.6\, \mathrm{GeV}$.

Figure \ref{fig:hypersurface2} shows the averaged temperature and baryon chemical potential of particlization hypersurface elements within specific rapidity ranges. 
For comparison, chemical freezeout parameters obtained from RHIC-BES data \cite{BES} are also shown. 
Up to $|\eta_{\mathrm{s}}| \approx 4$, averaged $\mu_{\mathrm{B}}$ are almost vanishing as expected.
    Then, $\mu_{\mathrm{B}}$ gradually increases with rapidity and reaches its maximum value in the range $7 \leqq |\eta_{\mathrm{s}}| \leqq 8$. 
Since the beam rapidity $y_{\mathrm{beam}}$ in this collision energy is $y_{\mathrm{beam}} \approx 8$, averaged $\mu_{\mathrm{B}}$ in $8 \leqq |\eta_{\mathrm{s}}| \leqq 9$ is smaller than that of $7 \leqq |\eta_{\mathrm{s}}| \leqq 8$ due to the baryon stopping. 
In comparison with the chemical freezeout parameters from RHIC-BES data, averaged $\mu_{\mathrm{B}}$ in $7 \leqq |\eta_{\mathrm{s}}| \leqq 8$ is comparable to that of Au+Au collisions at $\sqrt{s_{\mathrm{NN}}} = 7.7\, \mathrm{GeV}$.

\section{Summary}
We described the fluidization of baryon number in the entire system, from midrapidity to forward rapidity, of high-energy heavy-ion collisions using the extended DCCI2 model.
As a result, we found the existence of large fluctuations of baryon chemical potentials at midrapidity.
This means that regions with highly negative baryon chemical potential appear even at midrapidity. 
In forward rapidity, averaged baryon chemical potentials become as large as those of chemical freezeout parameters at RHIC-BES energies. 
This suggests that the high $\mu_{\mathrm{B}}$ region of the QCD phase diagram can be accessed via rapidity scan analysis in high-energy collisions, such as those at the top RHIC and LHC energies.

\section*{Acknowledgement}
The work by S.F. was supported by JST SPRING Grant No. JPMJSP2169. 
The work by T.H. was supported by JSPS KAKENHI Grant No. JP23K03395.


\begin{thebibliography}{00}

\bibitem{Li}
M.~Li and J.~I.~Kapusta, Phys.~Rev.~C \textbf{99}, 014906 (2019)

\bibitem{Kanakubo}
Y.~Kanakubo \textit{et al.}, Phys.~Rev.~C \textbf{105}, 024905 (2022)

\bibitem{pythia8}
T.~Sjöstrand \textit{et al.}, Comput.~Phys.~Commun. \textbf{191}, 159 (2015)

\bibitem{pythia8_angantyr}
C.~Bierlich \textit{et al.}, JHEP \textbf{1610}, 139 (2016)

\bibitem{NEOS}
A.~Monnai \textit{et al.}, Phys.~Rev.~C \textbf{100}, 024907 (2019)

\bibitem{BES}
L.~Adamczyk \textit{et al.} (STAR Collaboration), Phys.~Rev.~C \textbf{96}, 044904 (2017)

\end{thebibliography}
\end{document}